\documentclass[aps,a4paper,12pt]{revtex4}

\usepackage{float}
\usepackage{fancyhdr}
\usepackage{color}
\usepackage{graphicx}
\usepackage{epsfig} 
\usepackage{amssymb}
\usepackage{amsmath}
\usepackage{booktabs}
\usepackage{hyperref}
\usepackage[footnotesize]{caption}

\def\bra#1{\mathinner{\langle{#1}|}}
\def\ket#1{\mathinner{|{#1}\rangle}}

\def\Bra#1{\left<1>}
{\catcode`\|=\active\gdef\Braket#1{\left<\mathcode`\|"8000\let|\bravert {#1}\right>}}
\def\bravert{\egroup\,\vrule\,\bgroup}

\def\Tr{\mathop{\mbox{\normalfont Tr}}\nolimits}
\def\K{{\mathcal K}}
\begin{document}

\title{Entanglement of several blocks in fermionic chains.}

\author{F. Ares}
\affiliation{Departamento de F\'{\i}sica Te\'orica, Universidad de Zaragoza,
50009 Zaragoza, Spain}
\author{J. G. Esteve}
 \email{esteve@unizar.es}
 \affiliation{Departamento de F\'{\i}sica Te\'orica, Universidad de Zaragoza,
50009 Zaragoza, Spain}
\affiliation{Instituto de Biocomputaci\'on y F\'{\i}sica de Sistemas
Complejos (BIFI), 50009 Zaragoza, Spain}
  \author{F. Falceto\footnote{Corresponding author.} }
\email{ falceto@unizar.es}
 \affiliation{Departamento de F\'{\i}sica Te\'orica, Universidad de Zaragoza,
50009 Zaragoza, Spain}
\affiliation{Instituto de Biocomputaci\'on y F\'{\i}sica de Sistemas
Complejos (BIFI), 50009 Zaragoza, Spain}

%\date{\today; \quad Filename:\ {\tt\jobname.tex}}

%\begin{quote}
%{\tt  [Filename: \jobname.tex] }
%\end{quote}

\begin{abstract} 
In this paper we propose an expression for the entanglement entropy 
of several intervals in a stationary state of a free, translational 
invariant Hamiltonian in a fermionic chain. We check numerically the 
accuracy of our proposal and conjecture a new formula for the 
asymptotic behaviour of principal sub-matrices of a Toeplitz matrix.
\end{abstract}

\maketitle

\section{Introduction}

For the last years a considerable effort has been invested
to understand the entanglement of quantum systems.
This is, in Schr\" odinger words, {\it the characteristic trait of 
quantum mechanics} \cite{Schrodinger} and, as we now understand, 
the property that makes the
quantum computation to overtake the classical one.
Moreover, the study of the entanglement has a considerable 
interest from many other perspectives,
ranging from condensed matter physics \cite{Doyon,Latorre} or quantum field theory
\cite{Calabrese2009,CasiniJPA}
to black hole physics \cite{Solodukhin}
and the holographic principle \cite{Nishioka}. 

One of the preferred magnitudes to characterise entanglement 
is the R\'enyi entropy of the reduced state, that provides  
information on the full entanglement spectrum. Let
us consider a bipartite system such that
its Hilbert space can be written as the tensor
product $\mathcal{H}=\mathcal{H}_X\otimes \mathcal{H}_Y$,
of the Hilbert space of subsystems $X$
and $Y$ in which we have divided it. If $\rho$
is the density matrix that describes the state of
the whole system, the R\'enyi entropy of $X$
is defined as
$$S_\alpha(X)= \frac{1}{1-\alpha}\log\Tr(\rho_X^\alpha),$$
where $\rho_X=\Tr_Y(\rho)$ is the reduced density matrix of $X$ with
$\Tr_Y$ denoting the partial trace to the complementary subsystem $Y$.
In the limit $\alpha\to 1$, we obtain von Neumann entropy,
$$S_1(X)=-\Tr(\rho_X\log \rho_X).$$
If the system is in a pure state $\ket{\Psi}$, then
$\rho=\ket{\Psi}\bra{\Psi}$ and $S_\alpha(X)=S_\alpha(Y)$. In that
case, this quantity provides a very appropriate measure
of the degree of entanglement between $X$ and $Y$
in the state $\ket{\Psi}$ \cite{Plenio}. Furthermore, it encodes
universal properties of extended
systems in the neighbourhood of quantum critical points \cite{Doyon}.

The study of $S_\alpha(X)$ for fermionic chains is specially interesting
and simple because on one hand side, they
can be mapped to spin chains by means of a non local Jordan-Wigner transform
and, on the other side, we can apply both analytical techniques 
and efficient numerical algorithms. 
In this respect, much work has 
been done when $X$ is a single interval. In this case, a general result \cite{Ares}
for the eigenstates of a free, translational invariant
Hamiltonian can be obtained using the fact that the correlation matrix of an interval
is of the Toeplitz type (this property was first noticed by Jin
and Korepin \cite{Jin} for the ground state, and it is also
applied in e. g. \cite{Kadar, Eisler, Keating, Alba}). There are other different approaches. 
In particular, conformal field theory (CFT) is a powerful tool for the ground
state entanglement entropy when the chain is described by a local
and critical Hamiltonian \cite{Holzhey, Calabrese, Vidal} and it can be extended
to excited states too \cite{Sarandy, Alcaraz}. 
On the numerical side, 
we can reduce the complexity of computing the R\'enyi entropy
which, in principle, grows exponentially with the size of the subsystem
to a polynomial dependence.
This is possible thanks to the relationship obtained in \cite{Vidal,Peschel} 
between the density matrix and the two-point correlation functions for 
situations, like ours, in which Wick factorisation holds. 

A natural extension of the previous works is to consider
a subsystem $X$ composed of disjoint intervals. 
There are some recent papers where this problem is 
addressed for the ground state of a local Hamiltonian, which can be analysed 
using CFT \cite{Caraglio, Furukawa, Calabrese2, Facchi, Fagotti, Casini}.
If we try to apply the previous technique to this case 
we find that, although Wick factorisation still holds and therefore the
complexity of the computation grows polynomially, the corresponding matrix for 
several blocks is no longer of the Toeplitz type and the asymptotic expansion
for its determinant is not known in the literature, so far.

In this work, inspired by the previous analytical results and
some particular examples, we conjecture a
general asymptotic expression of $S_\alpha(X)$
for the eigenstates of a free, translational invariant fermionic Hamiltonian
when $X$ is composed by several disjoint blocks.
We check our hypothesis numerically 
and trace back its origin to a conjecture on the  
determinant of a principal sub-matrix of a Toeplitz matrix.

The paper is organised in
the following fashion. In the next section,
we introduce the notation and review the results for the entropy
of a single interval. In section III, we 
recall the results predicted by CFT for two disjoint intervals
in the ground state of local theories and we propose a new
conjecture for a general eigenstate. 
We check it numerically in section IV, while in section
V we generalise our formula for an arbitrary number of disjoint 
intervals and conjecture an asymptotic
expression for the determinant of a sub-matrix of a Toeplitz matrix.
Finally in section VI we collect our conclusions and possible 
continuations of our work.

\section{Entanglement entropy and Toeplitz determinants}\label{sec2}

We consider a chain of $N$ identical spin-less
fermions with $a_n$ and $a_n^\dagger$,
$n=1,\dots,N$, representing respectively the annihilation and creation  
operator for the site $n$. The only non vanishing 
anticommutation relations are
$$\{a_n,a_m^\dagger\}=\delta_{nm}.$$
Furthermore, we shall assume periodic 
boundary conditions: $a_{N+1}\equiv a_1$.

We consider the eigenstates of a free, translational invariant Hamiltonian,
\begin{equation}\label{state}
\vert \Psi_\mathcal{K} \rangle=  \prod_{k\in {\cal K}} b_k^\dagger \vert 0\rangle
\end{equation}
where b-operators are the discrete Fourier transform of $a$-operators,
$$b_k=\frac1{\sqrt{N}}\sum_{n=1}^N e^{\frac{2\pi i k n}{N}} a_n,\quad k=-N/2,\dots, N/2-1,$$
which also satisfy the canonical anticommutation relations. 
The ket $\vert 0\rangle$ represents the vacuum state
in the Fock space, ${\cal K}\subset\{-N/2,\dots, N/2-1\}$ is a particular
set of occupied modes
and  $$b_k^\dagger=\frac1{\sqrt{N}}\sum_{n=1}^N  e^{-\frac{2\pi i kn}{N}} a_n^\dagger$$
is the adjoint of $b_k$. 

We divide the chain into two subsets 
$X$ and $Y=\{1,\dots,N\}\setminus X$.
Adapted to this decomposition we can factor out the Hilbert 
space ${\cal H}={\mathcal{H}_X}\otimes{\mathcal{H}_Y}$. The goal is to
study the entanglement between these subsystems. 
 
In order to do that we must construct the reduced density matrix 
of each subsystem, that in general does not correspond to a pure state, and
compute its R\'enyi entropy. As it was discussed before, for pure states,
the entropy of a subsystem coincides with that of the complementary
one and provides a measurement for the entanglement between them.
Once we have obtained the reduced density matrix, we can compute
its R\'enyi entropy.
However, considering that the dimension of ${\cal H}_X$ is 
$2^{|X|}$, the computational time grows exponentially with the size $|X|$ 
of the subsystem.

Fortunately, there exists an algorithm \cite{Vidal,Peschel}, 
that can be applied in some cases and allows to reduce the 
exponential growth to a potential one. 
According to it, if the reduced density matrix
 satisfies the Wick decomposition property, i.e. the $n$-point functions 
factor out into two-point functions (see \cite{Peschel,Ares}),
the full reduced density matrix $\rho_X$, 
of dimension $2^{|X|}$, can be obtained from the two-point 
correlation matrix, whose dimension is $|X|$.
This dramatic gain of computational power allows us
to deal with larger subsystems $X$ 
without exhausting the computational capabilities. 
This is essential for us, as we will be interested in 
the asymptotic behaviour of the entanglement entropy for large
values of the size of the subsystem.

It is immediate to show that the Wick decomposition property, 
for the reduced density matrix,
follows from the same one for the full density matrix.
And the later enjoys this property for any pure state corresponding 
to a Slater determinant, like the one in (\ref{state}).

For these states the full density matrix, 
$\rho=\ket{\Psi_{\cal K}}\bra{\Psi_{\cal K}}$, 
preserves the total fermionic number and therefore 
$\Tr(\rho a_n a_m)=\Tr(\rho a^\dagger_n a^\dagger_m)=0$. 
Evidently,  this property is also fulfilled 
by the reduced density matrices.

\begin{figure}[H]
  \centering
    \resizebox{12cm}{4cm}{\includegraphics{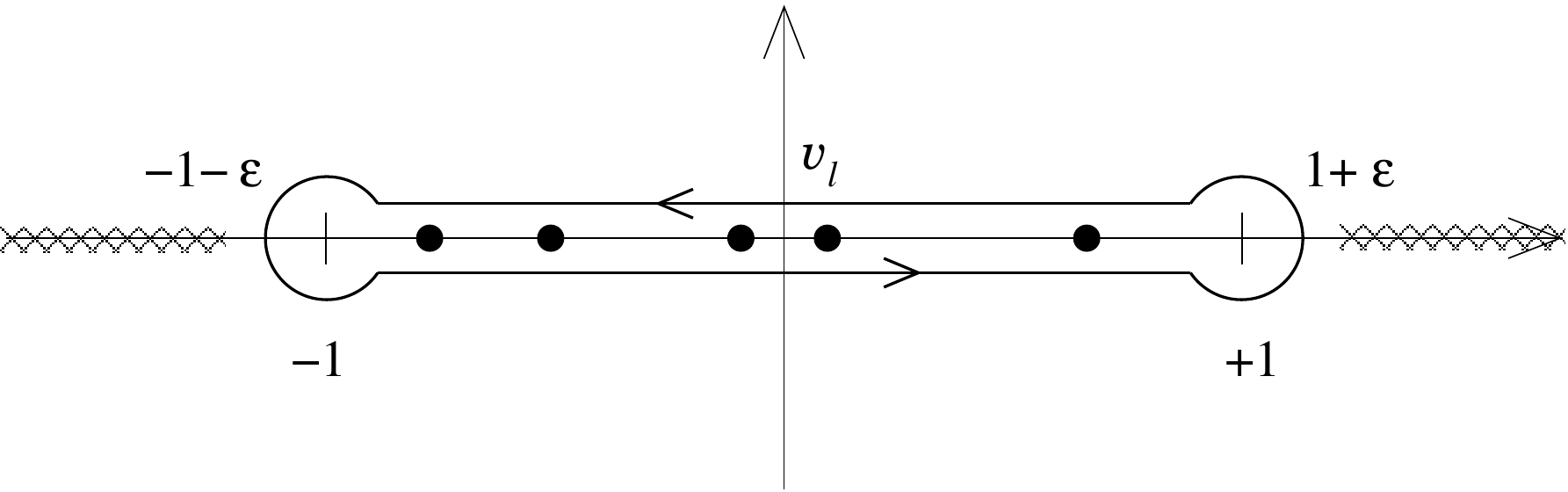}} 
    \caption{Contour of integration, cuts and poles for the computation of 
$S_\alpha(X)$. The cuts for the function $f_\alpha$ extend to $\pm\infty$.}
  \label{contorno0}
   \end{figure}

In that case it will be useful to introduce
the commutator expectation value matrix
\begin{equation}\label{correl}
(V_X)_{nm}=\Tr(\rho [a^\dagger_n, a_m]),
\quad n,m\in X
\end{equation}
in terms of which the R\'enyi entropy reads \cite{Ares}
\begin{eqnarray}\label{renyi1}
S_\alpha&=&\frac1{1-\alpha}\Tr\log\left[\left(\frac{I-V_X}2\right)^\alpha 
+ \left(\frac{I+V_X}2\right)^\alpha\right]\cr\cr
 &=&\lim_{\varepsilon
\to 0^+}\frac{1}{2\pi i}\oint_{\mathcal{C}} 
f_\alpha(1+\varepsilon,\lambda)\frac{d \log
\det(\lambda I-V_X)}{d\lambda}d\lambda.
\end{eqnarray}
In the second expression, we have made use of the Cauchy's 
residue theorem, with
$$f_\alpha(x,y)=\frac{1}{1-\alpha}
\log \left[\left(\frac{x+y}{2}\right)^\alpha+
\left(\frac{x-y}{2}\right)^\alpha\right],$$
and $\mathcal{C}$ the contour
depicted in the figure \ref{contorno0} that surrounds
all the poles of the logarithmic derivative of the 
determinant i.e. the eigenvalues, $v_l$, of $V_X$.

For the state $|\Psi_\mathcal{K}\rangle$,
the matrix $V_X$ can be written
\begin{equation}\label{vmatrix}
(V_X)_{n,m}=\frac1N\left(\sum_{k\in\mathcal{K}}e^{i\theta_k(n-m)}-
\sum_{k\not\in\mathcal{K}}e^{i\theta_k(n-m)}\right),
\end{equation}
with $\theta_k=2\pi k/N$.

We are interested in the large $N$ (thermodynamic) limit, 
in which case the role played by $\mathcal K$ is taken by a 
density $g(\theta)$, such that
\begin{equation}
\frac1{2\pi}\int_{-\pi}^\pi g(\theta)e^{i(n-m)\theta}d\theta
=
\lim_{N\to\infty} \frac1N\left(\sum_{k\in\mathcal{K}}e^{i\theta_k(n-m)}-
\sum_{k\not\in\mathcal{K}}e^{i\theta_k(n-m)}\right).
\end{equation}

A few examples, that will be useful in the following,
may help to understand the correspondence:

\begin{description}
\item[State 0:] The vacuum $|\Psi^{(0)}\rangle=|0\rangle$, that corresponds to
${\cal K}^{(0)}=\emptyset$ and has associated a constant density
$g^{(0)}(\theta)=-1$. 

\item[State 1:] $\K^{(1)}=\{-N/4+1,\dots,N/4-1,N/4\}$ which corresponds to 
$$
g^{(1)}(\theta)=
\begin{cases}
1&\mbox{for}\ \theta\in(-\pi/2,\pi/2],\\
-1&\mbox{for}\ \theta\not\in(-\pi/2,\pi/2],\\
\end{cases}
$$

\item[State 2:] $\K^{(2)}=\{-N/2+2,-N/2+4,\dots,0,2,\dots,N/2\}$ 
i.e. only even wave numbers are excited. The corresponding density
is also constant $g^{(2)}(\theta)=0$.

\item[State 3:] $\K^{(3)}=\{-N/4+2,-N/2+4,\dots,0,2,\dots,N/4\}$ 
i.e. even wave numbers between $-N/4+2$ and $N/4$ are excited. 
The corresponding density is 
$$
g^{(3)}(\theta)=
\begin{cases}
0&\mbox{for}\ \theta\in(-\pi/2,\pi/2],\\
-1&\mbox{for}\ \theta\not\in(-\pi/2,\pi/2],\\
\end{cases}
$$
\end{description}

\begin{figure}[H]
   \centering
   \includegraphics[width=.8\textwidth]{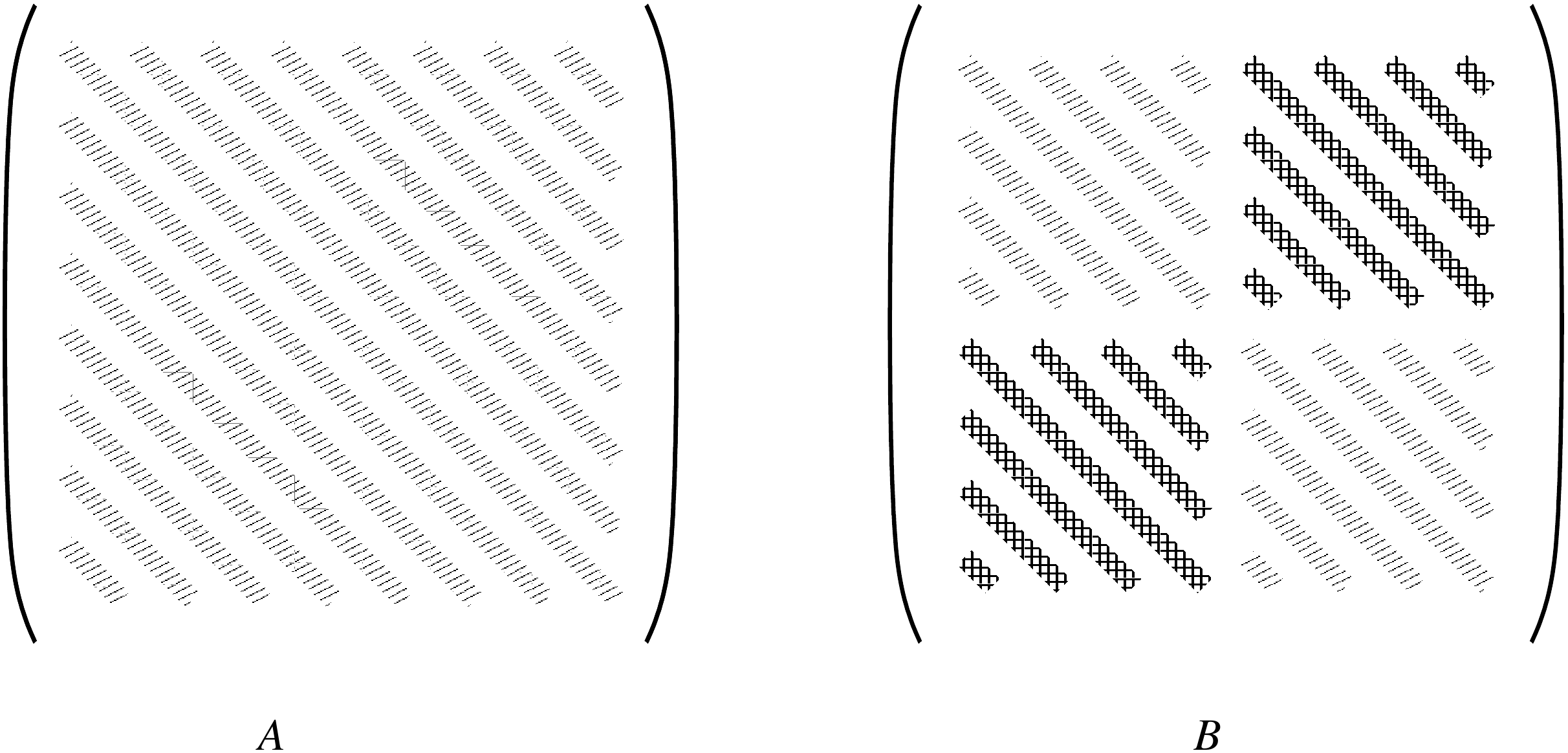}
   \caption{By the diagonal bars in the figure A, we represent
the defining property of a Toeplitz matrix: the entries along
any sub-diagonal parallel to the principal one are equal.
Figure B represents a block matrix in which each block is Toeplitz
but the full  matrix is not.}
  \label{toeplitz}\end{figure}

When subsystem $X$ is a single interval, i. e.  it is composed of 
consecutive sites, we have an extra property that allows us to compute the 
asymptotic behaviour of the R\'enyi entropy. 
In fact, in this case the matrix $V_X$ has all the entries of every 
sub-diagonal parallel to the main one equal, as it is represented in fig \ref{toeplitz} A. 
In other words, it is 
a diagonal-constant or Toeplitz matrix. Note that this property does not
hold, in general, if there is some gap between two sites in $X$; fig. \ref{toeplitz} B
provides an example of this.

It turns out that there exist asymptotic formulae
for computing the determinant of Toeplitz matrices and using them  in
(\ref{renyi1}) we can derive an expression for the entanglement entropy \cite{Jin, Ares}. 
Applying the general results to our case of interest 
and taking a piecewise constant $g(\theta)$ 
with discontinuities at $\theta_1,\dots, \theta_R$,
\begin{equation}\label{density}
g(\theta)=t_r,\quad \theta_{r-1}<\theta<\theta_r,
\end{equation} 
the entanglement entropy, when $X$ is an interval, reads
\begin{equation}\label{asymptotic}
S_\alpha(X)=A_\alpha |X|+B_\alpha\log|X|+C_\alpha+\dots
\end{equation}
where the dots represent terms that vanish in the large $|X|$ limit. 
The coefficients depend only on $g(\theta)$, not on $X$, and 
their computation is described below. 

In first place, the linear term is given by
\begin{equation}\label{A_alpha}
A_\alpha=\frac{1}{2\pi}\int_{-\pi}^{\pi} f_\alpha(1, g(\theta))d\theta.
\end{equation}

In order to compute $B_\alpha$ we introduce 
$$\omega_r(\lambda)=\frac{1}{2\pi}\log\left|\frac
{\lambda-t_r}{\lambda-t_{r-1}}
\right|,
$$
where the $t_r$'s are taken from (\ref{density}), and we define   
$$J_\alpha(r,r')=\frac{1}{2\pi}\int_{t_{r-1}}^{t_r} \frac{df_\alpha(1, \lambda)}{d\lambda} 
\omega_{r'}(\lambda) d\lambda, \quad r, r'=1,\dots, R.$$
Now 
\begin{equation}\label{B_alpha}
B_\alpha=2\sum_{r=1}^R J_ \alpha(r,r),
\end{equation}
where the upper limit in the sum is the number of 
discontinuity points in $g(\theta)$.

The constant term requires one more definition
\begin{equation}\label{I1}
 I_\alpha(r)=\frac{1}{2\pi i}\int_{t_{r-1}}^{t_r} \frac{d f_\alpha(1,\lambda)}{d\lambda}
 \log\left[\frac{\Gamma(1/2-i\omega_r(\lambda))}{\Gamma(1/2+i\omega_r(\lambda)}\right]d\lambda,
\end{equation}
where $\Gamma$ stands for the Gamma function.
From it we can write 
\begin{equation}\label{C_alpha}
C_\alpha=\sum_{r=1}^{R} I_ \alpha(r)
-
\sum_{1\leq r\not=r'\leq R}\log[2-2\cos(\theta_r-\theta_{r'})] J_ \alpha(r,r').
\end{equation}

The previous coefficients have been derived in \cite{Ares} using the 
Fisher-Hartwig conjecture for Toeplitz determinants, probed in our case by 
E. Basor \cite{Basor}. In \cite{Ares} we also perform the numeric computation of the entropy
and compare it with the asymptotic results to show the perfect agreement
between both calculations.

The results for the states defined before are the following:

\begin{description}

\item[State 0:] In this case $g^{(0)}(\theta)=-1$ and as we have $f_\alpha(1,-1)=0$
the linear coefficient vanishes. On the other hand, $g^{(0)}$  has no 
discontinuities  and therefore $B^{(0)}_\alpha$ and $C^{(0)}_\alpha$ vanish 
and the entanglement entropy $S^{(0)}_\alpha=0$, which can be obtained directly 
by noticing that  $|\Psi^{(0)}\rangle$ is separable.

\item[State 1:] As it is well known \cite{Jin} this state can be interpreted as
the ground state of a one dimensional,  local, critical theory.
Therefore, the results from conformal field theory apply and we should have \cite{Calabrese}
\begin{equation}\label{unoconformal}
S_\alpha^{\rm crit}(X)=\frac{c}6\frac{\alpha+1}\alpha\log|X|+ C_\alpha+\dots,
\end{equation}
where $c$ is the central charge of the underlying conformal field theory
and $C_\alpha$ a constant that depends on the details of the theory (non universal).\hfill\break
On the other hand, if we apply our general result and due to the fact that  $f_\alpha(1,\pm1)=0$ we have
$A^{(1)}_\alpha=0$. Also the logarithmic coefficient can be computed 
analytically to give
$B^{(1)}_\alpha=(\alpha+1)/(6\alpha)$, while the constant term does not have a
simple expression for general $\alpha$. Putting all together we have
$$S^{(1)}_\alpha(X)=\frac{\alpha+1}{6\alpha}\log |X| +C^{(1)}_\alpha+\dots$$
That agrees with (\ref{unoconformal}) for a central charge $c=1$.

\item[State 2:] If we write $|\Psi^{(2)}\rangle$
in the basis of {\it positions} we have
$$
|\Psi^{(2)}\rangle =\prod_{n=1}^{N/2}\frac1{\sqrt2}(a_n^\dagger+a^\dagger_{n+N/2})|0\rangle.
$$ 
Therefore it is easy to compute, exactly, the entanglement entropy for any subsystem.
In particular if $X$ is an interval of size smaller than $N/2$ the reduced density matrix is 
proportional to the identity
$$\rho_X=2^{-|X|} I$$
and therefore
\begin{equation}\label{entropy2}
S^{(2)}_\alpha(X)=|X|\log 2,
\end{equation}
which is independent of $\alpha$ and is the largest possible entropy for a mixed state
in a Hilbert space of dimension $2^{|X|}$.
If we derive  the coefficients of the expansion according to (\ref{A_alpha},\ref{B_alpha},\ref{C_alpha})
we get $A^{(2)}_\alpha=f_\alpha(1,0)=\log 2$ and, as we do not have any discontinuity,
$B^{(2)}_\alpha=C^{(2)}_\alpha=0$. Finally, the expansion leads to (\ref{entropy2}) that is exact 
in this case.

\item[State 3:] The entanglement entropy for this state combines the features  of the 
two previous ones: it has a non zero linear coefficient as $g(\theta)$ is different from $\pm1$ 
in some interval and it has discontinuities which give rise to the
logarithmic and constant coefficients. 
The linear term is easily computed to give
$A^{(2)}_\alpha=\log2/2$ and for integer $\alpha\geq 2$ we have \cite{Ares}, 
\begin{equation}\label{balpha}
B^{(2)}_\alpha=\frac{\alpha+1}{24\alpha}-\frac1{2\pi^2(\alpha-1)}
\sum_{l=1}^\alpha\left(\log\sin\frac{(2l-1)\pi}{2\alpha}\right)^2,\end{equation}
while for $\alpha=1$
$$B_1=\frac1{8}-\frac1{2}\left(\frac{\log 2}{\pi}\right)^2.$$
\end{description}

In this paper we want to go one step further and discuss the case when 
$X$ is made out
of several intervals. Our goal is to derive asymptotic formulae
for the entanglement entropy similar to (\ref{asymptotic}). In this case, 
however,
we can not use the Fisher-Hartwig formula and we will approach 
the problem by using the results from conformal field theory and
performing numeric computations. Later on we will be 
able to derive a generalisation of Fisher-Hartwig conjecture that covers the
case of several intervals.

\section{Two disjoint intervals}

As it was mentioned before, if the set $X$ is composed of non contiguous sites, 
for instance the union of two separate intervals $X=X_1\cup X_2$
with $X_1$ and $X_2$ made out of contiguous sites, the matrix $V_X$ is not of 
the Toeplitz type any more. This is shown pictorially in
Fig. \ref{toeplitz} B, where we represent the fact that while all four sub-matrices 
are Toeplitz, the full matrix is not: except for the main diagonal all the 
others have two kinds of entries.

In this situation we can not apply the Fisher-Hartwig conjecture and 
we should try to get insights on the behaviour of the entanglement entropy 
from a different source.

If the state we consider corresponds to the ground state of a critical, 
local, one dimensional system (like it happens for the state 1) we can use
the conformal invariance of the theory, that follows from the absence of a 
fundamental length (zero mass gap). This powerful symmetry
determine to some extent the behaviour of the entanglement entropy.

In this case \cite{Calabrese2} if we take $X=[u_1,v_1]\cup[u_2,v_2]$ with $u_1<v_1<u_2<v_2$,  global
conformal invariance leads to
\begin{equation}\label{dosconforme}
\Tr\rho_X^\alpha\sim K_\alpha \left(\frac{(v_1-u_1)(v_2-u_2)(v_2-u_1)(u_2-v_1)}{(u_2-u_1)(v_2-v_1)}\right)^{4\alpha\Delta_\alpha}
{\cal G}(y)
\end{equation}
where 
\begin{equation}\label{invariant}
y= \frac{(u_2-v_1)(v_2-u_1)}{(u_2-u_1)(v_2-v_1)}
\end{equation}
is the cross ratio $(u_1,v_1;u_2,v_2)$ which is invariant 
under the linear fractional transformations $z\mapsto (az-b)/(d-cz)$, 
${\cal G}$ is a non universal function
that depends on the details of the theory (see \cite{Calabrese2}),
$$\Delta_\alpha=\frac{c}{24}\left(1-\frac1{\alpha^2}\right)$$ 
is the conformal dimension of the insertions and, finally, $K_\alpha$ is a constant that 
will be fixed below.

In \cite{Casini} it is shown that for the ground state of critical 
free fermions (as it is our case) ${\cal G}=1$. Then if we compute 
the R\'enyi entropy we obtain
\begin{equation}\label{entrconforme}
S_\alpha(X)=\frac{c}6\frac{\alpha+1}\alpha
\log\frac{(v_1-u_1)(v_2-u_2)(v_2-u_1)(u_2-v_1)}{(u_2-u_1)(v_2-v_1)}+
\frac1{1-\alpha}\log K_\alpha+\dots.
\end{equation}
In order to determine the constant on the right we can take the limit of
large separation between $X_1$ and $X_2$ in which case the entropy should go to
the sum of the entropy of every interval. Therefore 
the constant on the right
of (\ref{entrconforme}) should be twice the constant for a single interval
$C_\alpha$ that we determined in the previous section.
 
In summary, for a critical, local theory for which $g(\theta)=\pm1$,
like the state 1, we should have
\begin{equation}
S_\alpha^{(1)}(X)=B_\alpha^{(1)}\log\frac{(v_1-u_1)(v_2-u_2)(v_2-u_1)(u_2-v_1)}{(u_2-u_1)(v_2-v_1)}+2C_\alpha^{(1)}+\dots,
\end{equation}
where the constants are those determined in the previous section for a single interval.
We have checked the expression against numerical results and it completely agrees.

Another insight on the problem can be gained by considering the state 2.
In this case, we can compute the entropy exactly and, provided
$|v_2-u_1|<N/2$ we have that the reduced 
density matrix is again proportional to the identity and
$$S_\alpha^{(2)}(X)= |X|\log 2.$$
where, in this case, $|X|=|v_1-u_1|+|v_2-u_2|$. It should be noticed that 
the coefficient that multiplies the size of the subsystem
coincides with that for a single interval.

If we put together the two previous results we conjecture the following expression for
the R\'enyi entanglement entropy of two disjoint intervals for a general state:
\begin{equation}\label{entrdos}
S_\alpha(X)=A_\alpha|X|+B_\alpha\log\frac{(v_1-u_1)(v_2-u_2)(v_2-u_1)(u_2-v_1)}{(u_2-u_1)(v_2-v_1)}+2C_\alpha+\dots,
\end{equation}
where the coefficients $A_\alpha, B_\alpha,C_\alpha$ are those determined in (\ref{A_alpha},\ref{B_alpha},\ref{C_alpha})
for a single interval.
The expression (\ref{entrdos}) should be valid in the thermodynamic limit, while the dots stand for contributions that vanish when 
$|v_i-u_j|\to\infty$ for $i,j=1,2$.

In the next section we will investigate numerically the 
validity of our conjecture.

\section{Numerical results}

In order to check the accuracy of the previous expression it will be useful 
to introduce the so called mutual information $I_\alpha(X_1,X_2)$ defined by
$$I_\alpha(X_1,X_2)=S_\alpha(X_1)+S_\alpha(X_2)-S_\alpha(X).$$
From (\ref{entrdos}) we obtain the simple expression
\begin{equation}\label{information}
I_\alpha(X_1,X_2)=-B_\alpha\log y,
\end{equation}
where $y$ is the cross  ratio in Eq. (\ref{invariant}).
The logarithmic coefficient $B_\alpha$ is obtained from the Fisher-Hartwig conjecture
for the expansion of the entropy for a single interval,
as it is discussed in section II.

We perform numerical calculations for the state
3 choosing an infinite chain, $N\to\infty$
when $\alpha=1$ (von Neumann entropy) and $\alpha=2$.
Since the Wick decomposition is satisfied, they can be carried out 
diagonalising the correlation matrix $V_X$ (\ref{vmatrix})
for this configuration and then applying the first 
equality of (\ref{renyi1}). 
As we have discussed before,
this implies an impressive simplification ($V_X$
has dimension $|X|$ while $\rho_X$, $2^{|X|}$) which
we need to explore the asymptotic behaviour of the mutual
information. In fact, in the following numerical computation we have 
covered values of $|X|$ from $100$ to
$5500$. Notice that these values would be  
absolutely out of reach in a direct computation
using $\rho_X$, with dimension $2^{5500}$.
 
For the diagonalisation, we have employed the corresponding
routine for real symmetric matrices included in the
GNU Scientific Library \cite{Galassi} for C, which works in double
precision. 

\begin{figure}[H]
   \centering
   \includegraphics[width=1.1\textwidth]{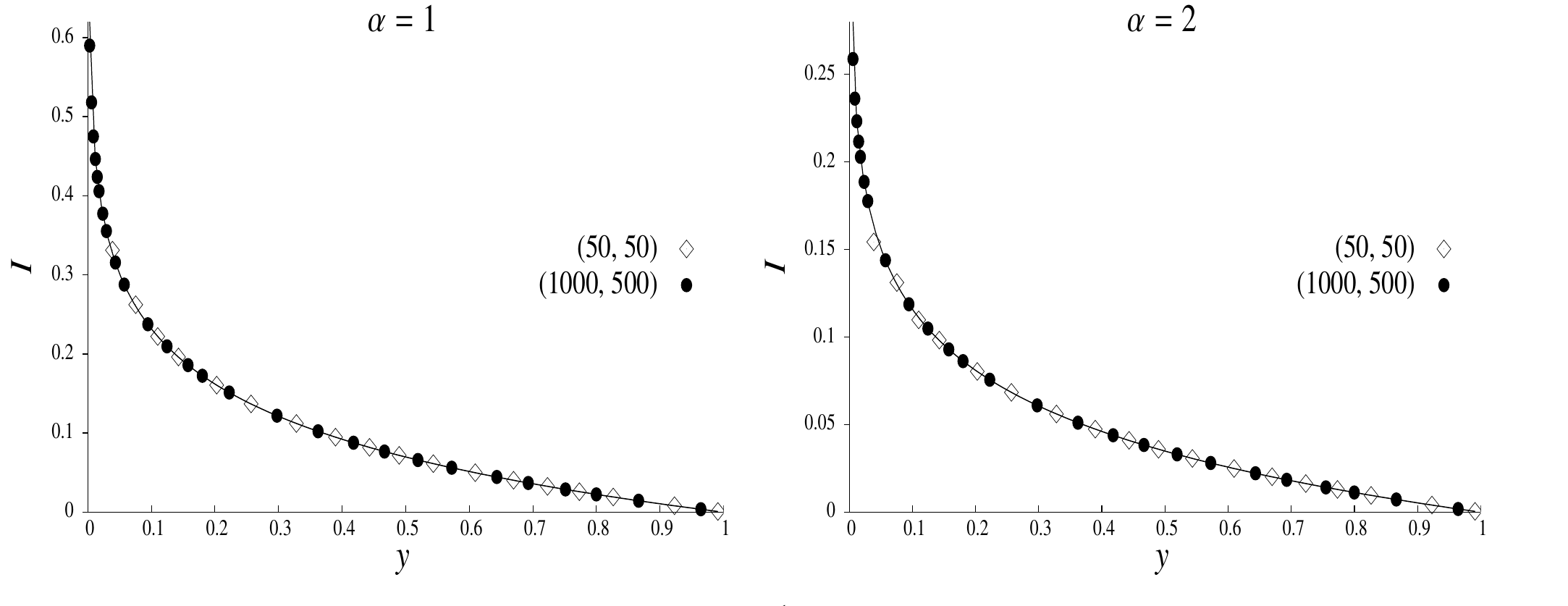}
   \caption{Two blocks mutual information when $\alpha=1$
   (left panel) and $\alpha=2$ (right panel) as a function of $y$ for 
   the state 3. With $\diamond$ we represent the numerical
   value for two blocks made out of 50 sites each of them,
   varying their separation from 1 up to 500 sites. The $\bullet$
   corresponds to two blocks of lengths 1000 and 500 sites, 
   separated each other between 1 and 1000 sites. The
   continuous line depicts the function (\ref{information})
   with the $B_\alpha$ evaluated from the Fisher-Hartwig conjecture.}
  \label{two}\end{figure}

Dots in figure \ref{two} represent the numerical
results for two different sizes of the blocks. 
The continuous line stands for the analytic candidate
(\ref{information}). Remember for state 3 $B_\alpha$ is given
by the expression (\ref{balpha}) for an integer $\alpha\geq 2$. 
For $\alpha=2$ it leads,
$$B_2^{(3)}=\frac{1}{16}-\frac{1}{4}\left(\frac{\log 2}{\pi}\right)^2=0.050330...$$
Notice that for this configuration, we have
$$B_1^{(3)}=2B_2^{(3)}=0.100660...$$

There is an excellent agreement between the numerical results and the 
analytical expression (\ref{information}) we have proposed. From the plots 
it is also apparent that the mutual information only depends on
the cross ratio of the involved distances. We have also
performed the computations with other block lengths and different states,
finding the same accordance with the theoretical prediction.

\section{Several intervals and a generalisation of 
the Fisher-Hartwig conjecture}

The results obtained in the previous section for two intervals
can be immediately generalised to the case of $p$ disjoint intervals. 
Namely, consider
$$X=\bigcup_{i=1}^p[u_i,v_i],\quad u_i<v_i<u_{i+1}$$
then, keeping in mind the results of conformal field theory 
and (\ref{entrdos}) it is natural to write
\begin{equation}\label{entrn}
S_\alpha(X)=A_\alpha\sum_{i=1}^p(v_i-u_i)+
B_\alpha\log\frac{\prod_{i,j=1}^p|u_i-v_j|}{\prod_{j>i} (u_j-u_i)(v_j-v_i)}
+ pC_\alpha,
\end{equation}
where the last two terms are taken directly from the conformal field
expressions and the first one reflects the extensivity of the linear term.
Like before, this expression should be  valid in the thermodynamic limit,  
and the dots represent terms that vanish when $|u_i-v_j|\to\infty,\ i,j=1,\dots,p$.

In order to check (\ref{entrn}) it will be useful to introduce the analogue of the mutual 
information for $p$ intervals, given by
\begin{equation}\label{pmutual}
I_\alpha([u_1,v_1],\dots,[u_p,v_p])=\sum_{i=1}^pS_\alpha([u_i,v_i])-
S_\alpha(\bigcup_{1=1}^p [u_i,v_i]).
\end{equation}
Note that this is different from the tripartite mutual information of \cite{CasiniJHEP}.
Actually according to our results the latter 
vanishes in the asymptotic limit.

If (\ref{entrn}) is correct we should have
\begin{equation}\label{Iconjectured}
I_\alpha([u_1,v_1],\dots,[u_p,v_p])=-B_\alpha\log \prod_{i<j} y_{ij}
\end{equation}
where
$$y_{ij}=\frac{(u_j-v_i)(v_j-u_i)}{(u_j-u_i)(v_j-v_i)}$$
is the cross ratio of $(u_i,v_i;u_j,v_j)$.

\begin{figure}[H]
   \centering
   \includegraphics[width=1.1\textwidth]{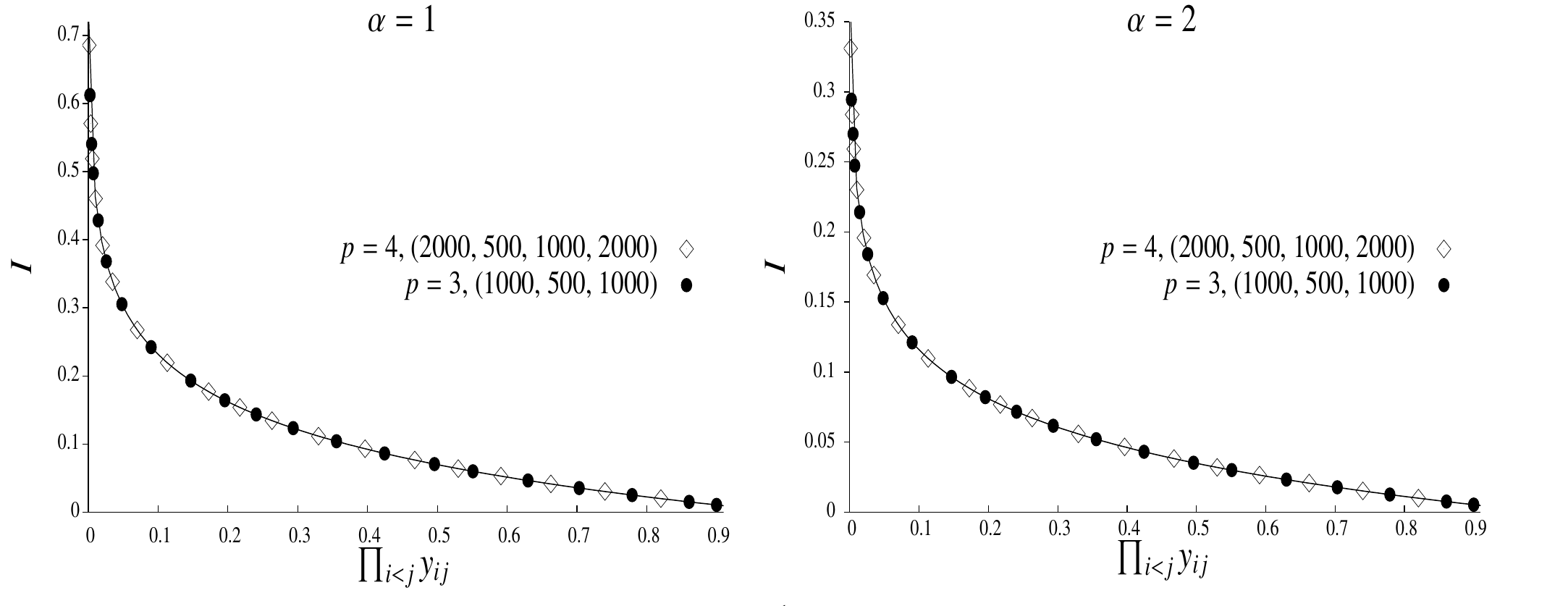}
   \caption{Mutual information, for the state 3, of three ($\bullet$)
   and four ($\diamond$) blocks as a function of the product of
   the possible cross ratios $y_{ij}$. For $p=3$, we have chosen two intervals
   of 1000 sites and one of 500 sites which is separated from
   one of the former by 1500 sites, while the remaining distance
   is modified from 1 up to 99000 sites. For $p=4$, we take intervals
   of lengths 2000, 500, 1000, 2000. The distance between the first couple
   is 1500 sites, the break between the smallest blocks is also fixed,
    5000 sites, whereas the another one is increased between 1 and 99000
    sites.}
  \label{threefour}\end{figure}

In order to verify (\ref{Iconjectured})
we have computed  (\ref{pmutual}) numerically for 
three and four intervals for the state
3 with different sizes and distances for the intervals. 
The results are shown in fig. (\ref{threefour})
together with the conjectured behaviour stated in
(\ref{Iconjectured}). The agreement is extraordinary.

The expression for the entropy in Eq. (\ref{entrn}) can also be written as
a combination of that for single intervals,  
\begin{equation}\label{entrnentr}
S_\alpha(X)=\sum_{i\geq j}S_\alpha([u_j,v_i])+
\sum_{i < j}\big( S_\alpha([v_i,u_j])- S_\alpha([v_i,v_j])-
S_\alpha([u_i,u_j]) \big).
\end{equation}
It is not difficult 
to show that combining the expression above and 
(\ref{asymptotic}) we derive Eq. (\ref{entrn}). 
However (\ref{entrnentr}) has the virtue of showing more clearly the 
possible origin of our result for several intervals,
as we will see below.
An expression similar to (\ref{entrnentr}) for the ground state of local 
critical theories has been derived in
\cite{Hubeny} applying the holographic principle. See also
ref. \cite{CasiniJHEP}.

In the computation of the entropy for a single interval a key step was
to use the asymptotic expansion of the Toeplitz determinant 
which appears in the integrand of (\ref{renyi1}) and is well
known in the literature. However, as it was noticed before, for more than 
one interval the correlation matrix is not Toeplitz any more but it is a
principal sub-matrix of a Toeplitz matrix, as it is depicted in fig. \ref{toeplitz} B.

On the other hand, as it is made explicit in (\ref{renyi1}), 
$S_\alpha(X)$ depends linearly on the logarithm of the determinant of this 
sub-matrix. 
Therefore, the relation (\ref{entrnentr}) can be derived from an
analogous property for the determinants of principal
sub-matrices of a Toeplitz matrix. 

In order to formulate the conjecture, consider a general Toeplitz matrix $T$
with piecewise smooth symbol $g(\theta)$ and for any set of indices $K$ define
$D(K)=\det(T_{nm}),\quad n,m\in K$. Then, the property for the determinant
of the principal sub-matrix that we hypothesise can be stated as follows.

\begin{figure}[H]
   \centering
   \includegraphics[width=.9\textwidth]{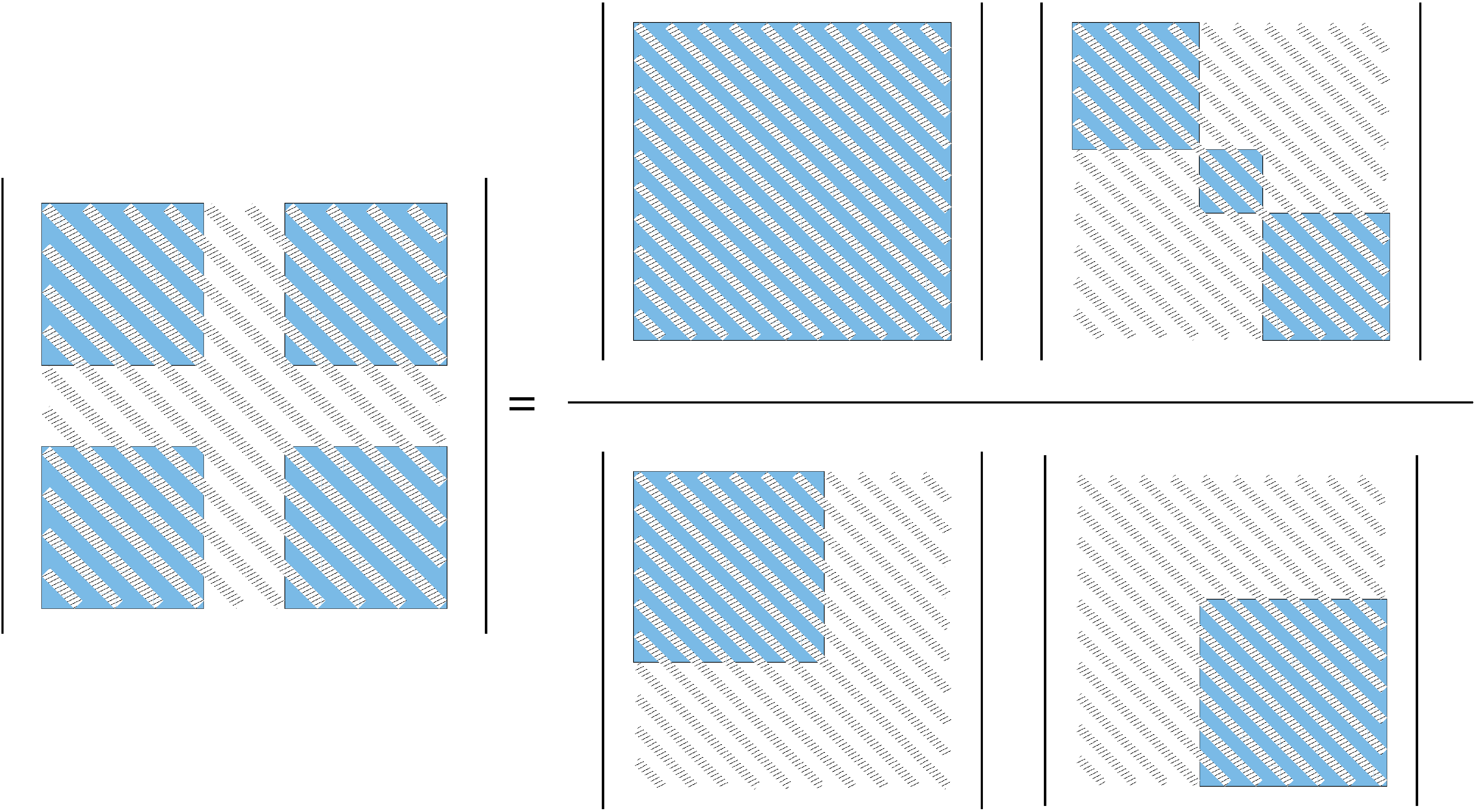}
   \caption{Graphical representation of the conjecture (\ref{toeplitzn})
for $p=2$. In the left hand side we represent the determinant of the 
shadowed sub-matrix that, in general,
is not Toeplitz. 
In the right hand side, however, the determinants of the shadowed sub-matrices
are of the Toeplitz type (or product of these).} 
  \label{conjecture}\end{figure}

{\bf Conjecture:} 
\begin{equation}\label{toeplitzn}
D(\bigcup_{i=1}^p [u_i,v_i] )\simeq\prod_i D([u_i,v_i])
\prod_{i< j} \frac{D([u_i,v_j]) D([v_i,u_j])}
{D([u_i,u_j])D([v_i,v_j])},
\end{equation}
where $\simeq$ stands for the equality
of the asymptotic behaviour when
$|v_i-u_j|\to\infty$, for $i,j=1,\dots, p$.
Notice that all determinants on the right 
hand side are of the Toeplitz type and therefore,
using the Fisher-Hartwig conjecture, 
(\ref{toeplitzn}) allows the computation of
the scaling of general principal sub-matrices of a 
Toeplitz matrix. 
Formula (\ref{toeplitzn}) is depicted graphically, for $p=2$, in fig. \ref{conjecture}.

Of course, the previous conjecture has been checked indirectly when 
he have computed
the entropy of several intervals, but we think it is worth studying 
for a more general, piecewise smooth symbol. We choose the following 
$$
g(\theta)=
\begin{cases}
\frac{1}{4}\left(3+\sin\theta\right),&\theta\in(-\pi,0]\\
\frac{1}{4}\left(3+\cos\theta\right),&\theta\in(0,\pi],
\end{cases}
$$
from which it is easy to compute the corresponding matrix:
$$
T_{nm}=\frac1{2\pi}\int_{-\pi}^\pi g(\theta)e^{i(n-m)\theta}d\theta=
\begin{cases}
0,& \mbox{for}\ n-m\ \mbox{odd},\\
\displaystyle\frac{1}{4\pi}\frac{(n-m)i+1}{(n-m)^2-1}+\frac34\delta_{nm},& 
\mbox{for}\ n-m\ \mbox{even}.\\
\end{cases}
$$

We check (\ref{toeplitzn}) in this particular case
for $p=2$, studying the analogue of the two blocks mutual information
for determinants,
\begin{equation}\label{analogue}
I_D([u_1, v_1]\cup[u_2, v_2])=\log D([u_1, v_1])+\log D([u_2, v_2])-
\log(D[u_1, v_1]\cup[u_2, v_2])
\end{equation}

\begin{figure}[H]
   \centering
   \includegraphics[width=1.1\textwidth]{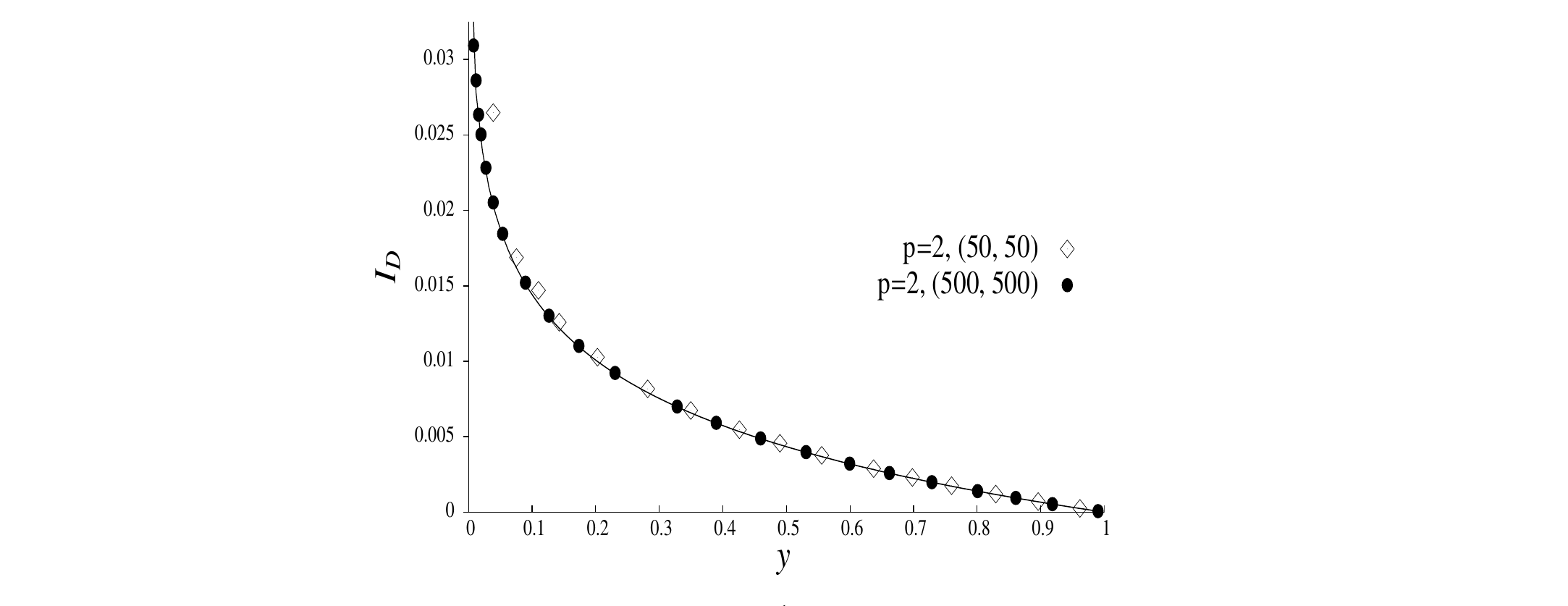}
   \caption{Analogue of mutual information for determinants, (\ref{analogue}),
against the cross ratio of $(u_1,v_1;u_2,v_2)$. 
The $\diamond$  represents the numerical results for two subsets of size $|u_1-v_1|=|u_2-v_2|=50$ 
while the gap between them, $|v_1-u_2|$, varies between 1 and 200. 
The $\bullet$ corresponds to two subsets of length 500 separated
by a distance between 1 and 4500. 
The continous line is the conjectured analytical expression of (\ref{inf_det}).}
  \label{det}\end{figure}

Then, applying the Fisher-Hartwig conjecture for the Toeplitz sub-matrices which appears
in (\ref{toeplitzn}), we should have
\begin{equation}\label{inf_det}I_D([u_1, v_1]\cup [u_2, v_2])=-B_D \log y,\end{equation}
where the coefficient $B_D$ can be obtained analytically from the general 
expression (\ref{B_alpha}). 
For this particular case it is
$$B_D=\frac1{2\pi^2}\left[\int_{3/4}^1 \frac1{\lambda}\log\left|\frac{\lambda-1}{\lambda-3/4}\right|d\lambda+
\int_{1/2}^{3/4} \frac1{\lambda}\log\left|\frac{\lambda-3/4}{\lambda-1/2}\right|d\lambda\right]=0.0062607...$$
In figure \ref{det}, we represent by dots the numerical value of $I_D$ while
the solid line represents the logarithmic dependence (\ref{inf_det}) predicted by our conjecture, with the coefficient computed above.

It is certainly remarkable the agreement between our conjecture and the 
numerical results. Due to the asymptotic nature of our formulae,
the accordance with the numerical result should be poorer when the separation
between the intervals is only of a few sites. This is especially striking
when we study the determinant of two small subsets as we can see in figure \ref{det}. 

\section{Conclusions}

In this paper we have studied the entanglement between
two subsystems made out of several disjoint intervals
for the  eigenstates of a uni-dimensional
fermionic chain described by a free, translational
invariant Hamiltonian.

Similarly to the case of a single interval, since these
states satisfy the Wick decomposition, we can compute the R\'enyi entanglement 
entropy employing the two-point correlation matrices restricted to
one of the subsystems. This reduces the complexity which, in principle,
grows exponentially with the size of our subsystem. On the other hand, for two or more
disjoint blocks, these correlations are no more Toeplitz matrices (in fact, they
are block matrices where each block is Toeplitz). Hence, the Fisher-Hartwig 
expansion does not hold.

Therefore, we have been forced to resort to a different strategy in order to 
gain understanding on the behaviour of the entropy for several intervals.
One of the sources for our intuition comes from conformal field theory, 
that can be applied to the ground state of local, gap-less theories. 
The other source is the opposite: non local theories (ladders)
with a mass gap. In both cases there is an alternative way of computing 
the entropy:
using conformal invariance in the first instance and  
by direct computation in the second case.

Extending these partial results, we propose
a general asymptotic expansion for an arbitrary translational invariant
state. We have checked that it perfectly matches with the numerical
value for different states and several block numbers and sizes.

Finally, from this result we can conjecture the solution for a more 
general problem: the asymptotic behaviour of 
the determinant of general principal sub-matrices of a Toeplitz matrix.
Our result relates the determinant of our sub-matrix to the product
of several others of the Toeplitz type, which 
combined with the Fisher-Hartwig theorem, provides an asymptotic scaling for 
this kind of determinants. We have numerically verified our conjecture
for a particular Toeplitz matrix with a piecewise smooth symbol.

One of the motivations for working with a chain of spin-less fermions
is its relation with spin chains. In the general case, however, the resulting
Hamiltonian for the fermionic chain, although it is still quadratic, does not
preserve the total fermionic number. Then, the two point function involves more
coefficients and the resulting matrix is of the, so called, block Toeplitz type.
It would be nice to extend our results for general stationary states
to these systems.

\noindent{\bf Acknowledgments:} Research partially supported by grants 2012-E24/2,  
DGIID-DGA and FPA2012-35453, MINECO (Spain).


\begin{thebibliography}{XXX}

\bibitem{Schrodinger} E. Schr\"odinger
\textit{Discussion of Probability Relations Between Separated Systems},
Proc. Camb. Philos. Soc., 31, 555â€“563 (1935); 32,  446â€“451 (1935).

\bibitem{Doyon} P. Calabrese, J. Cardy, B. Doyon, 
\textit{Entanglement entropy in extended quantum systems}, 
J. Phys. A: Math. Theor. 42, 500301 (2009); 
arXiv:0708.2978v1[cond-mat.stat-mech]

\bibitem{Latorre} J. I. Latorre, A. Riera,
\textit{A short review on entanglement in quantum spin systems},
J. Phys. A: Math. Theor. 42,  504002  (2009);
arXiv:0906.1499 

\bibitem{Calabrese2009} P. Calabrese, J. Cardy,
\textit{Entanglement entropy and conformal field theory},
J. Phys. A: Math. Theor. 42, 504005 (2009); 
arXiv:0905.4013, 

\bibitem{CasiniJPA}  H. Casini, M. Huerta,
\textit{Entanglement entropy in free quantum field theory},
J.Phys. A: math. Theor. 42, 504007 (2009);
arXiv:0905.2562, 

\bibitem{Solodukhin} S. N. Solodukhin
\textit{Entanglement entropy of black holes},
Living Rev. Relativity 14, (2011), 8.
arXiv:1104.3712, 

\bibitem{Nishioka} T. Nishioka, S. Ryu, T. Takayanagi,
\textit{Holographic Entanglement Entropy: An Overview},
J. Phys. A: Math. Theor. 42, 504008 (2009);
arXiv:0905.0932

\bibitem{Plenio} M. B. Plenio, S. Virmani, 
\textit{An introduction to entanglement measures}, 
Quantum Information \& Computation 7, 1-51 (2007); 
arXiv:quant-ph/0504163v3 


\bibitem{Ares} F. Ares, J. G. Esteve, F. Falceto, E. S\'anchez-Burillo, 
\textit{Excited states entanglement in homogeneous fermionic chains}, 
J. Phys A: Math Theor. to appear.
arXiv: 1401.5922[quant-phys]

\bibitem{Jin} B. Q. Jin, V. E. Korepin, 
\textit{Quantum spin chain, Toeplitz determinants and the Fisher-Hartwig conjecture}, 
J. Stat. Phys. 116, 79 (2004); 
arXiv:quant-ph/0304108v4

\bibitem{Kadar} Z. K\'adar, Z. Zimbor\'as, 
\textit{Entanglement entropy in quantum spin chains with broken reflection symmetry}, 
Phys. Rev. A 82, 032334 (2010), 
arXiv:1004.3112 [quant-ph]

\bibitem{Eisler} V. Eisler, Z. Zimbor\'as, 
\textit{ Area law violation for the mutual information in a nonequilibrium steady state}, 
arXiv:1311.3327[cond-mat.stat-mech]

\bibitem{Keating} J. P. Keating, F. Mezzadri, 
\textit{Entanglement in quantum spin chains, symmetry classes of random matrices and CFT}, 
Phys. Rev. Lett. 94 050501 (2005);
arXiv:quant-ph/0504179

\bibitem{Alba} V. Alba, M. Fagotti, P. Calabrese, \textit{Entanglement entropy of excited states}, 
J. Stat. Mech. 0910:P10020 (2009); 
arXiv:0909.1999v2[cond-mat.stat.mech]

\bibitem{Holzhey} C. Holzhey, F. Larsen, F. Wilczek, \textit{Geometric and renormalized entropy in conformal field theory}, 
Nucl. Phys. B 424, 443-467 (1994); 
arXiv:hep-th/9403108v1 

\bibitem{Calabrese} P. Calabrese, J. Cardy, 
\textit{Entanglement entropy and Quantum Field Theory}, 
J. Stat. Mech. 0406:P06002 (2004); 
arXiv:hep-th/0405152v3 

\bibitem{Vidal} G. Vidal, J.I. Latorre, E. Rico, A. Kitaev, 
\textit{Entanglement in quantum critical phenomena}, 
Phys. Rev. Lett. 90, 22: 227902-227906 (2003); 
arXiv:quant-ph/0211074v1

\bibitem{Sarandy} F. C. Alcaraz, M. S. Sarandy, 
\textit{Finite size corrections to entanglement in quantum critical systems}, 
Phys. Rev. A 78, 032319 (2008); 
arXiv:0808.0020v2[quant-ph]

\bibitem{Alcaraz} F. C. Alcaraz, M. Ib\'a\~nez Berganza, G. Sierra, 
\textit{Entanglement of low-energy excitations in CFT}, 
Phys. Rev. Lett. 106, 201601 (2011); 
arXiv:1101.2881v2[cond-mat.stat-mech];\hfill\break
\textit{Entanglement of excited states in critical spin chains}, 
J. Stat. Mech. P01016 (2012); 
arXiv:1109.5673v1[cond-mat.stat-mech]

\bibitem{Peschel} I. Peschel, 
\textit{Calculation of reduced density matrices from correlation functions}, 
J. Phys. A: Math. Gen. 36, L205 (2003); 
arXiv:cond-mat/0212631v1 (2002)                  

\bibitem{Caraglio} M. Caraglio, F. Gliozzi, 
\textit{Entanglement entropy and twist fields}, 
JHEP, 11(2008)076 ; 
arXiv:0808.4094v2[hep-th]

\bibitem{Furukawa} S. Furukawa, V. Pasquier, J. Shiraishi, 
\textit{Mutual information and boson radius in a c=1 critical system in one dimension}, 
Phys. Rev. Lett. 102, 170602 (2009);
arXiv:0809.5113v3[cond-mat-stat-mech]

\bibitem{Calabrese2} P. Calabrese, J. Cardy, E. Tonni, 
\textit{Entanglement entropy of two disjoint intervals in conformal field theory}, 
J. Stat. Mech. P11001 (2009); 
arXiv:0905.2069v2[hep-th]

\bibitem{Facchi} P. Facchi, G. Florio, C. Invernazzi, S. Pascazio, 
\textit{Entanglement of two blocks of spins in the critical Ising model}, 
Phys. Rev. A 78, 052302 (2008);
arXiv:0808.0600v2[quant-phys]

\bibitem {Fagotti} M. Fagotti, 
\textit{New insights into the entanglement of disjoint blocks}, 
Europhys. Lett. 97, 17007 (2012); 
arXiv:1110.3770v2[cond-mat-stat-mech]

\bibitem{Casini} H. Casini, M. Huerta, 
\textit{Reduced density matrix and internal dynamics for multicomponent regions},
Class. Quant. Grav. 26 185005 (2009); 
arXiv:0903.5254v2[hep-th]                  

\bibitem{Fisher} M. E. Fisher, R. E. Hartwig, 
\textit{Toeplitz determinants, some applications, theorems and conjectures},
Adv. Chem. Phys. 15, 333-353 (1968)

\bibitem{Basor} E. L. Basor, 
\textit{A localization theorem for Toeplitz determinants}, 
Indiana Math. J. 28, 975-983 (1979)

\bibitem{BasorMorrison} E. L. Basor, K. E. Morrison, 
\textit{The Fisher-Hartwig conjecture and Toeplitz eigenvalues}, 
Linear Algebra and its applications, 202, 129-142 (1994)

\bibitem{Galassi} M. Galassi et al, 
\textit{GNU Scientific Library Reference Manual},
(3rd Ed.), ISBN 0954612078.
\texttt{http://www.gnu.org/software/gsl/}

\bibitem{CasiniJHEP} H. Casini, M. Huerta,
\textit{Remarks on the entanglement entropy for disconnected regions},
JHEP 03(2009)048;
arXiv: 0812.1773v2[hep-th]

\bibitem{Hubeny} V. Hubeny, M. Rangamani,
\textit{Holographic entanglement entropy for disconnected regions},
JHEP 03(2008)006;
arXiv: 0711.4118v2[hep-th]

\end{thebibliography}
\end{document}